\input psfig.sty
\documentstyle[preprint,aps,floats]{revtex}

\newcommand{\beq}{\begin{equation}}
\newcommand{\eeq}{\end{equation}}
\newcommand{\beqa}{\begin{eqnarray}}
\newcommand{\eeqa}{\end{eqnarray}}
\newcommand{\asl}{$a_{SL}$}
\newcommand{\apks}{$a_{\psi K_S}$}
\newcommand{\app}{$a_{\pi\pi}$}

\newcommand{\bbar}{$B^0-\bar B^0$}
\newcommand{\goff}{$\Gamma_{12}$}
\newcommand{\moff}{$M_{12}$}
\def\sm{Standard Model}
\newcommand{\mot}{$M_{12}$}
\newcommand{\msmot}{$M^0_{12}$}
\newcommand{\dmot}{$\delta M_{12}$}
\def\epem{$e^+e^-$\ }
\newcommand{\betam}{$\tilde\beta$}

\newcommand{\vtd}{$V_{td}$}

\newcommand{\fb}{$\sqrt{B_B}f_B$}

\def\npb#1{Nucl.\ Phys.\ {\bf B #1}}
\def\plb#1{Phys.\ Lett.\ {\bf B #1}}
\def\prd#1{Phys.\ Rev.\ {\bf D #1}}
\def\prl#1{Phys.\ Rev.\ Lett. {\bf #1}}

\def\rmp{Rev.\ Mod.\ Phys.\ }

\def\epj#1{Eur.\ Phys.\ J.\ {\bf #1}}

\draft
\begin{document}
{\tighten


\preprint{
\vbox{\hbox{LBNL-43159}
      \hbox{UCB-PTH-99/18}
      \hbox{hep-ph/9904480}
      \hbox{April 1999} }}

\title{
Constraining the CKM Parameters using CP Violation in semi-leptonic 
$B$ Decays} 

\author{
Robert N. Cahn
}
\address{
\vbox{\vskip 0.true cm}
Theoretical Physics Group \\
Earnest Orlando Lawrence Berkeley National Laboratory \\
University of California, Berkeley, California 94720
}
\medskip

\author{
Mihir P. Worah
}
\address{
\vbox{\vskip 0.true cm}
Department of Physics \\
University of California, Berkeley, California 94720 \\
{\rm and} \\ 
Theoretical Physics Group \\
Earnest Orlando Lawrence Berkeley National Laboratory \\
University of California, Berkeley, California 94720
}
\maketitle

\begin{abstract}%
We discuss the usefulness of 
the CP violating semi-leptonic asymmetry \asl\ not only as a signal of 
new physics, but also as a tool in constraining the CKM parameters.
We show that this technique could yield useful results in the first
years of running at the $B$ factories.
We present the analysis graphically 
in terms of \moff, the dispersive part of the \bbar\ mixing
amplitude. This is complementary to the 
usual unitarity triangle representation and often allows a cleaner
interpretation of the data. 

\end{abstract}


}
\renewcommand{\thefootnote}{\alph{footnote}}

\newpage


The goal of the B physics programs soon to begin at \epem, $ep$ and
$p{\overline p}$ colliders around the world is to test the Standard
Model's predictions for CP violation. It is important to have a means
of quantifying these tests.  
One seeks measurements and analyses that would not only offer clean
signals of new physics, but would also allow the extraction of
fundamental \sm\ parameters. 

A reasonable assumption is that the new physics that affects the
\bbar\ mixing amplitude does not affect either the 
$B$ meson decay amplitudes or CKM unitarity.%
\footnote{For a general analysis of the case where the
decay amplitudes are also affected, see \cite{GW}.}
In that case, one can couple the already measured 
values of $|V_{ub}|$ and $\Delta m_B$ 
with the measurements of \apks\ and \app, the CP violating asymmetries 
in the decays $B\to \psi K_S$ and $B \to \pi\pi$ respectively, 
to construct the unitarity triangle 
and also disentangle the new physics contributions to \bbar\ mixing from
the \sm\ ones \cite{GNW}. 
A drawback of this approach is that the unitarity triangle analysis 
tends to mix up the experimental errors, which are often
quite small, with the theoretical errors that arise in relating these
measurements to CKM parameters.  
An attractive alternative is to focus on the dispersive part of the 
off-diagonal matrix element, $M_{12}$, of the \bbar\ mixing matrix 
\cite{WS,GOTO}. In this construction, the data is graphically
represented in the complex \moff\ plane \cite{GOTO}. 
An advantage of this representation is a 
separation between the experimental uncertainty in $\Delta m_B$ from the 
theoretical uncertainty in its calculation.
A shortcoming of both approaches is that discrete ambiguities in
relating \apks\ and \app\ to CKM phases leads to multiple solutions
for the \sm\ and new physics parameters \cite{GNW,GQ}. 
Thus, one needs additional
information to try and resolve these.

In this paper we use the graphical representation in the \moff\ plane
to highlight the information that can
be obtained from a measurement of \asl, the CP violation in semi-leptonic
$B$ decays. 
The sensitivity of \asl\ to new physics is well known \cite{SX,RS}. 
We show, in addition, how one can use constraints on, or the observation
of \asl\ to restrict allowed regions in the \sm\ parameter space.
Such an analysis requires a precise calculation of 
$\Delta\Gamma$, the \bbar\ width difference. This calculation uses the
notion of local quark-hadron duality, and moreover depends on certain
non-perturbative ``bag factors''. 
We propose tests of its consistency, and note that  
its precision should be significantly improved in the near future
by new input from lattice calculations. 


Under the assumption that the $B$ decay amplitudes are not affected,
all the new physics effects can be expressed in terms of one complex
number: the new contribution to the dispersive part of the 
\bbar\ mixing amplitude, $M_{12}$. Explicitly, we write 
\beq
M_{12}=M_{12}^0+\delta M_{12}\label{m12s}
\label{dm}
\eeq
where $M_{12}^0$ represents the Standard Model contribution and 
$\delta M_{12}$ is a complex number representing the new physics
contribution. Also useful is the equivalent representation \cite{GNW},
\beq
M_{12} = r^2 e^{i2\theta}M_{12}^0
\label{m12_def}
\eeq
We will work in the convention where the phase of $M_{12}^0$ 
is $2\beta$, thus that of \moff\ is 
$2(\beta+\theta) \equiv 2\tilde\beta$. 
(Note, that these phases are 
measured relative to that of the $b \to c \bar c d$ decay amplitude).

The magnitude of $M_{12}$ is well determined:
\beq
|M_{12}|=\Delta m_B/2
\eeq
where  $\Delta m_B= 0.470\pm 0.019$ ps$^{-1}$ =$3.09\times 10^{-13}$
 GeV\ \cite{PDGA}. 
We can use this to represent the actual value of \mot\ = \msmot\ + \dmot\ 
as lying somewhere on the unit circle centered at the origin of the
complex \moff\ plane (where
all data are rescaled by the experimentally determined central value of
$\Delta m_B/2$). The phase of \moff, $2\tilde\beta$, 
will be obtained from the CP asymmetry in $B \to \psi K_S$:
\beq
a_{\psi K_S} = \sin 2\tilde\beta.
\eeq

\begin{figure}
\begin{center}
\psfig{file=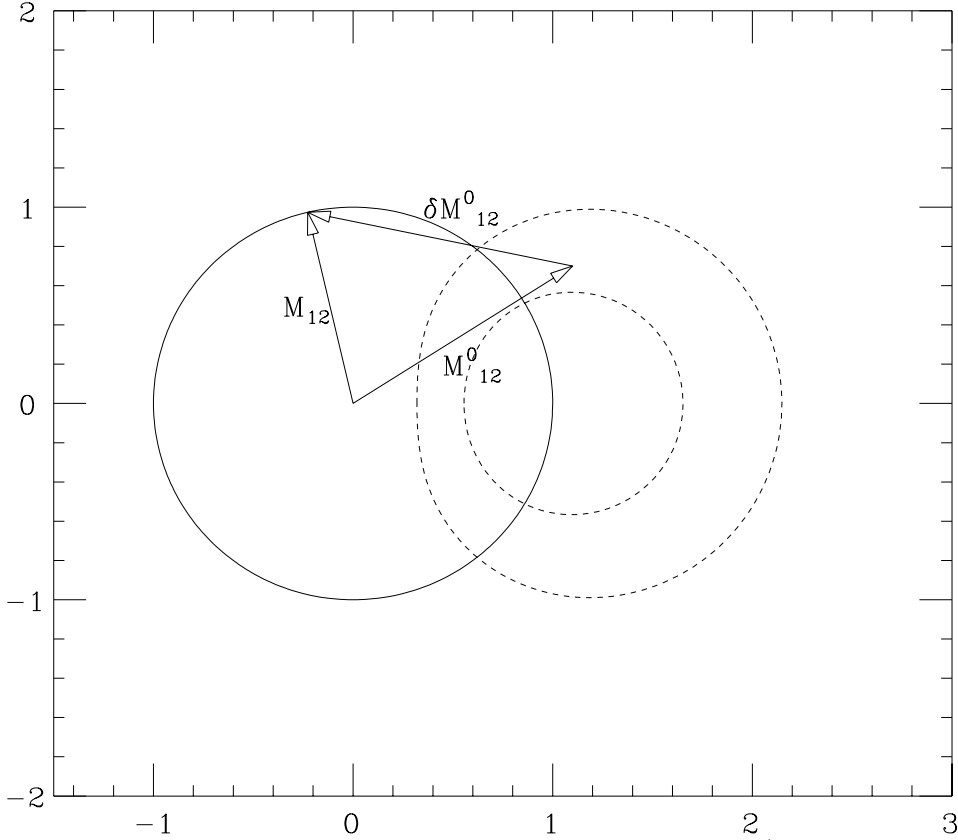,width=5.0in,angle=90}
\begin{tighten}{
\caption{\small{The complex $M_{12}$ plane shown in units of 
$\Delta m_B/2$.
The measured value of $M_{12}$ is thus a thin annulus. The Standard
Model
contribution, $M_{12}^0$, falls within the distorted annulus.  The
shape is determined by the values of \vtd\ allowed by unitarity,
given the measured terms in the CKM matrix.  The central value for
$B_Bf_B^2=(200~{\rm MeV})^2$ is used here.  The total off-diagonal
matrix element is the sum of the Standard Model contribution and the
new physics: $M_{12}=M^0_{12}+\delta M_{12}$.}}
\label{m12plane}
}\end{tighten}
\end{center}
\end{figure}

We can plot the allowed \sm\ region in this plane using \cite{BBH}
\beq
M_{12}^0 =
\frac{G_F^2}{12\pi^2}m_Bm_t^2\eta_BB_Bf_B^2(V_{tb}V_{td}^{*})^2
S_0(x_t).
\label{m0}
\eeq
Here $S_0(x_t) \simeq 0.784x_t^{0.76}$ 
(this is a fit to the exact formula \cite{BBH}) 
is a kinematical factor with
$x_t=m_t^2/m_W^2$. The factor $\eta_B=0.55$ is a QCD correction, and
typical values for $\sqrt{B_B}f_B$ are $200\pm 40$ MeV. Using
$m_t=165$ GeV, we find
\beq
M_{12}^0 = \frac{\Delta m_B}{2}\left | \frac{V_{tb}V_{td}^*}{0.0086}
\right |^2 \left ( \frac{\sqrt{B_B}f_B}{200~{\rm MeV}} \right )^2
e^{2i\beta}
\label{m120_1}
\eeq
In the absence of new physics, $M_{12}^0=M_{12}$ and one can directly
use $\Delta m_B$ to infer a value for $|V_{tb}V_{td}^*|$.  Although
this is not possible if new physics is present, we can still use the 
unitarity of the CKM matrix to plot an allowed region for the 
\sm, and thus constrain $|V_{tb}V_{td}^*|$. Using 
\beq
{V_{ub}\over V_{cb}}=ae^{-i\gamma},\qquad 0.06 \leq a\leq 0.10
\eeq
and considering $V_{ud}= 0.975$, $V_{cd} = -0.220$, and 
$V_{cb}=0.0395$ \cite{PDGC}
as well determined relative to the other uncertainties in the
problem, we obtain
\beqa
|V_{tb}V_{td}^*|e^{-i\beta} &=&-(V_{cb}V_{cd}^*+V_{ub}V_{ud}^*)
\nonumber \\
&=& -0.0395(-0.220 + 0.975ae^{-i\gamma}).
\label{m120_2}
\eeqa
Using this relation in Eq. (\ref{m120_1}), we find that
as $a$ covers the stated range and $\gamma$ varies over $0$
to $2\pi$, \msmot\ covers a region of the complex $M_{12}$ plane
as shown in Fig.~\ref{m12plane}.
\moff, the full \bbar\ mixing amplitude can lie anywhere on the solid
circle, and $M_{12}^0$, the \sm\ contribution lies somewhere in the
region between the two dashed curves. If there were no new physics,
\moff\ would have to lie on the solid circle in one of the two regions
where it intersects with the allowed \sm\ area.

\begin{figure}
\psfig{file=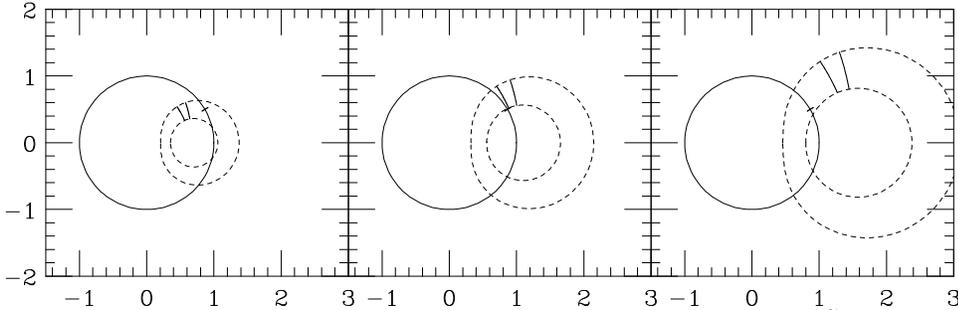,width=5.0in,angle=90}
\begin{tighten}
\caption[a]{\small{The complex $M_{12}$, in units of
$\Delta m_B/2$.  The value of $\tilde\beta=0.262$ is indicated by the tick
mark on the unit circle.  The allowed range of $\gamma$ derived from
$\sin 2(\gamma+\tilde\beta)=0.43\pm 0.20$ is a slice of the annular
region.  The three figures correspond, left to right, to the values
$\sqrt{B_B}f_B= 160, 200, 240$ MeV.}}
\label{262}
\end{tighten}
\end{figure}

Measuring \app, the CP asymmetry in
$B\to \pi\pi$ would give
$\sin 2(\gamma+\tilde\beta)$ (once the penguin effects are determined)
Since, in principle, both \betam\ and $\gamma+\tilde\beta$ are known,
$\gamma$ itself is known.  For fixed $\gamma$ the allowed
region for \msmot\ is a curve extending from the inner to the
outer boundary of the \msmot, as shown in Fig. ~\ref{262}.

\begin{figure}
\psfig{file=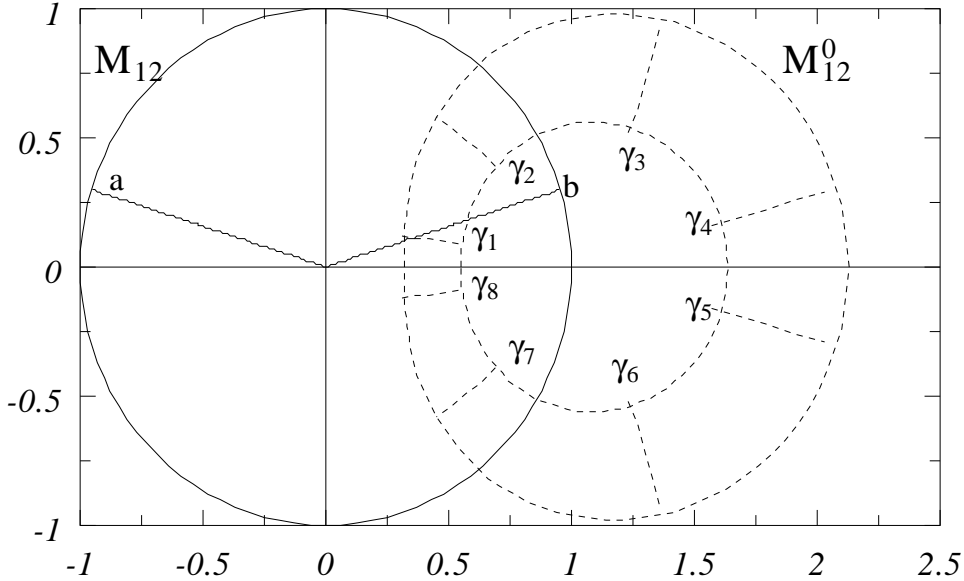,width=5.0in}
\vskip 0.2in
\begin{tighten}{
\caption[a]{\small{
The complex $M_{12}$ plane in units of $\Delta m_B/2$.
We show the two-fold discrete ambiguity in the value of $2\tilde\beta$
(the points $a$ and $b$)
and the eight-fold ambiguity in $\gamma$ 
(the curves labeled $\gamma_1 ... \gamma_8$)
resulting from the
measurements \apks = 0.3 and \app = -0.7.
We have used $\sqrt{B_B}f_B= 200$ MeV in obtaining the Standard Model
region.}}
\label{gamma_8}
}\end{tighten}
\end{figure}

Thus, in principle, the CP violating measurements \apks\ and \app\
allow us to disentangle the \sm\ contribution to \bbar\ mixing from the 
new physics contribution. In this construction \dmot\ is the vector
extending from the $\gamma$ curve in the allowed region for \msmot\ to 
the tip  of the \betam\ vector on the unit circle. This procedure is
complementary to that used in \cite{GNW} to obtain $r$ and $\theta$
from these measurements. The advantage in this case is a 
clean separation of
the experimental uncertainties in $\Delta m_B$ which are small, from
the theory errors in the \sm\ contribution to it. 
Just as in the unitarity triangle analysis, however, discrete ambiguities in 
obtaining the phases \betam\ amd $\gamma$ lead to multiple allowed
regions, thus muddying the situation \cite{GNW,GQ}. 
Without additional inputs, the measurements of \apks\ and \app\ only
allow us to extract $2\tilde\beta$ up to a two-fold ambiguity, and $\gamma$
up to an eight-fold ambiguity. We illustrate
this in Fig. \ref{gamma_8} based on perfect measurements of the
quantities $a_{\psi K_S}=0.3$ and
$a_{\pi\pi}=-0.7$. As shown in the figure, the true value of the
\bbar\ mixing amplitude, $M_{12}$ could be either of the points
labeled $a$ or $b$. The \sm\ contribution to it, $M_{12}^0$ could lie
on any one of the curves labeled $\gamma_1$ through $\gamma_8$. 
If there is no new physics, 
one can use information from $K-\bar K$ mixing as well
as the fact that $\alpha$, $\beta$, and $\gamma$ are the angles of a
triangle to reduce these ambiguities to a simple two-fold ambiguity in
$\gamma$. This is not possible in the presence of new physics, and one
needs additional information in order to extract the \sm\ parameters
from the CP violating measurements \apks\ and \app.
Note, that the \apks\ value chosen here already tells us that there is new
physics present in the \bbar\ mixing amplitude. This can be seen from
the fact that neither of the point $a$ and $b$ on the 
\moff\ circle lies within the allowed \sm\ region. 
Although there exist techniques that allow a direct extraction of the
angle $\gamma$, these are either experimentally difficult \cite{GLW},
or suffer from theoretical uncertainties and sensitivity to new
physics \cite{FM}.
We will now
discuss how a measurement of, or constraints on \asl\ 
restricts the allowed \sm\ parameter space and helps 
resolve discrete ambiguities.
%


In machines running at the $\Upsilon(4s)$, \asl\ is
measured by the asymmetry in dilepton events with same sign leptons
coming from both $B$ decays:
\beq
a_{SL} \equiv \frac{N(l^+l^+)-N(l^-l^-)}{N(l^+l^+)+N(l^-l^-)}
\eeq
where $N(l^+l^+)$ [$N(l^-l^-)$] defines the number of times
a $B^0-\bar B^0$ pair decays into a pair of positively [negatively]
charged leptons.The source of $a_{SL}$ is CP violation in
the $B^0-\bar B^0$ mixing matrix and it arises due to a phase between the
absorbtive and dispersive parts of the \bbar\ mixing amplitude,
\beq
a_{SL} = {\rm Im}(\frac{\Gamma_{12}}{M_{12}})
       = \left |\frac{\Gamma_{12}}{M_{12}} \right |\sin\phi_{12}
\label{a_sl_def}
\eeq
where $\phi_{12}$ is the phase between $\Gamma_{12}$ and $M_{12}$.
In the Standard Model, $a_{SL}$ is unobservably small,
$ \sim 10^{-3}$ because
$|\frac{\Gamma_{12}}{M_{12}}| \sim 10^{-2}$ and because the GIM mechanism
results in $\sin\phi_{12} \sim m_c^2/m_b^2 \sim 10^{-1}$.
Thus, new physics can enhance $a_{SL}$ by increasing
$|\frac{\Gamma_{12}}{M_{12}}|$ and/or $\sin\phi_{12}$.

In order for new physics to significantly affect \goff, one would 
need either large new
decay amplitudes into known states that are common to both $B^0$
and $\bar B^0$, or to introduce additional, exotic common final
states. Such a scenario 
could enhance both the factors mentioned above, and could lead to
$a_{SL} \sim 0.1$ \cite{GPW}. 
This would be detected in the very early stages of
data taking at the asymmetric $B$ factories, with only 
about $10^6$ \bbar\ pairs. 
Here we concentrate on the more likely possibility 
where the new heavy particles contribute
to \moff\ but not \goff. This could lead to enhancements of
$\sin\phi_{12}$, thus allowing $a_{SL} \sim 0.01$ \cite{SX,RS}, 
which would be observable in about one year of running at the $B$ factories.

\begin{figure}
\psfig{file=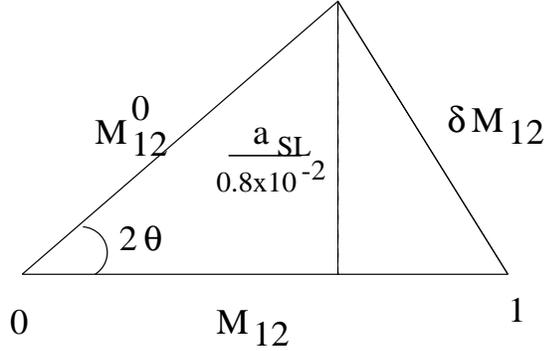,width=5.0in}
\vskip 0.2in
\begin{tighten}
\caption[a]{\small{
The relationship between $M_{12}$, $M_{12}^0$, and $a_{SL}$.
The perpendicular distance between $M_{12}$ and the tip of the
$M_{12}^0$ vector is given by $a_{SL}/0.8\times 10^{-2}$. Where
$0.8\times 10^{-2}$ is the calculated central value of
$\Gamma_{12}^0/M_{12}^0$.  
}}
\label{triangle}
\end{tighten}
\end{figure}

Within the \sm, at leading order we have \cite{Hagelin,BSS,BBD}
\beqa
\Gamma_{12}^0 &= {\displaystyle{
-\frac{G_F^2m_b^2m_B}{24\pi}[
\frac{5}{3}\frac{m_B^2}{(m_b+m_d)^2}
     (K_1-K_2)f_B^2B_S(V_{tb}V_{td}^{*})^2}} \nonumber \\ 
& {\displaystyle{
+ \frac{8}{3}(K_1+\frac{K_2}{2})f_B^2B_B(V_{tb}V_{td}^{*})^2 + 
8(K_1+K_2)f_B^2B_B\frac{m_c^2}{m_b^2}V_{cb}V_{cd}^{*}V_{tb}V_{td}^{*}}}].
\label{g0}
\eeqa
Here $K_1=-0.39$ and $K_2=1.25$ \cite{BBD} 
are combinations of Wilson coefficients. $B_S$ and $B_B$ are the 
bag factors corresponding to the matrix elements of the operators 
$Q_S \equiv (\bar b d)_{S-P}(\bar b d)_{S-P}$ and
$Q \equiv (\bar b d)_{V-A}(\bar b d)_{V-A}$. Combining Eqs. (\ref{g0}) 
and (\ref{m0}), and using $m_b=4.5$ GeV, we have%
{\footnote{Note that obtaining the numerical result requires using
$\eta_B=0.88$ in Eq. (\ref{m0}) due to the different definition of
$B_B$ in Eq. (\ref{g0}). See Ref. \cite{BBD} for details.}}
\beq
\frac{\Gamma_{12}^0}{M_{12}^0} = -5.0\times 10^{-3}
                                 \left (1.4\frac{B_S}{B_B} + 0.24 +
                                 2.5 \frac{m_c^2}{m_b^2}
                                 \frac{V_{cb}V_{cd}^*}{V_{tb}V_{td}^*}
                                 \right ).
\label{sm_asl}
\eeq
In the vacuum saturation
approximation one has $B_S/B_B=1$ at some typical hadronic scale, and
this expectation is confirmed by a leading order lattice calculation
\cite{Pierro}. Although corrections to the vacuum saturation
value are unknown, a more precise lattice calculation of this ratio
should be available soon. 
This would result in a more reliable central
value with well defined errors 
(which are expected to be $\lesssim {\cal O}(25\%)$) \cite{Pierro}. 
Note, however, that the uncertainty due to the ratio of bag factors is
restricted to ${\rm Re}(\Gamma_{12}/M_{12}) \simeq \Delta\Gamma/\Delta 
m$, and that 
${\rm Im}(\Gamma_{12}/M_{12})$ which arises from the 
third term in the parenthesis
does not suffer from this uncertainty. Thus, \asl\ is precisely
calculated in the \sm.
From the measured value of $|V_{ub}/V_{cb}|$ and 
CKM unitarity we know that $|\sin\beta| < 0.35$.
Then, using $m_c^2/m_b^2=0.085$ 
and ${\rm Im}(V_{cb}V_{cd}^*/V_{tb}V_{td}^*) \sim \sin\beta$ 
leads to the limit
$a_{SL}^{SM} < 10^{-3}$ which is unobservably
small. To simplify matters, we will ignore this small phase in the
\sm\ value of \goff/\moff.
One can then write
\beqa
\frac{\Gamma_{12}}{M_{12}} &= & \frac{\Gamma_{12}}{M_{12}^0}
                              \frac{M_{12}^0}{M_{12}} \nonumber \\
                     &=& -(0.8\pm0.2) \times 10^{-2}\frac{e^{-i2\theta}}{r^2}
\label{g_by_m}
\eeqa
where we have used Eq. (\ref{m12_def}) and $B_S/B=1\pm0.25$ in 
Eq. (\ref{sm_asl}). 
Thus, Eqs. (\ref{a_sl_def}) and (\ref{g_by_m}) lead to
\beqa
a_{SL}& = & (0.8 \pm 0.2) \times 10^{-2}
          {\rm Im}(\frac{M_{12}^0}{M_{12}}) \nonumber \\
      &=& (0.8\pm 0.2) \times 10^{-2} \frac{\sin 2\theta}{r^2}
\label{asl}
\eeqa
Combining Eqs. (\ref{m12_def}) and (\ref{asl}) one sees that 
$M_{12}^0$ is given by a vector
at an angle $2\theta$ from $M_{12}$ and whose tip is a perpendicular 
distance $a_{SL}/0.8\times10^{-2}$ from it.
In Fig ~\ref{triangle} we demonstrate this
relation between \moff, \msmot, and \asl. 

\begin{figure}
\psfig{file=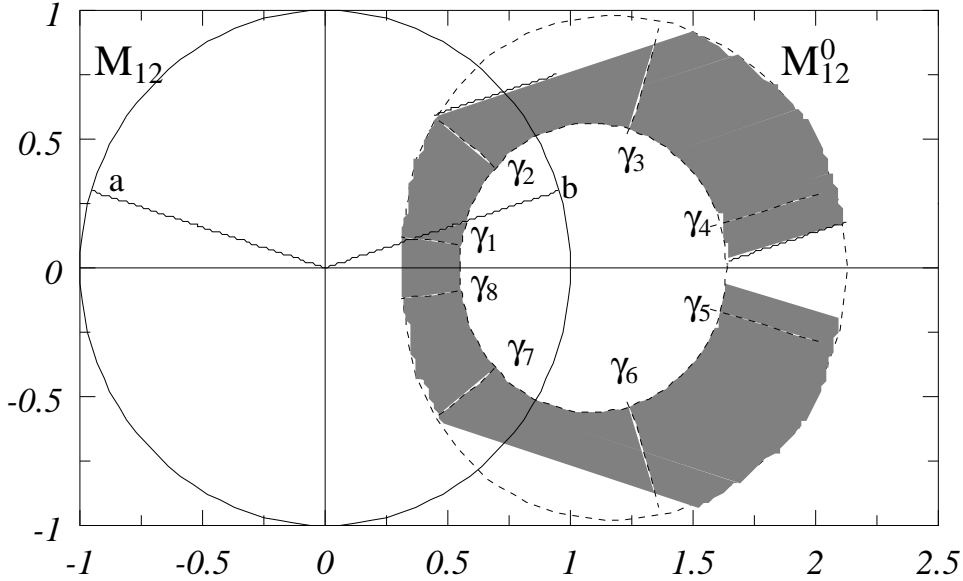,width=5.0in}
\vskip 0.2in
\begin{tighten}
\caption[a]{\small{
The complex $M_{12}$ plane in units of $\Delta m_B/2$.
The points $a$ and $b$ and the curves $\gamma_1...\gamma_8$ result
from the measurements \apks = 0.3 and \app = -0.7.
The shaded region corresponds to the allowed Standard Model parameter
space coming from a measurement of $|a_{SL}|<5 \times 10^{-3}$.
We have used $\sqrt{B_B}f_B= 200$ MeV in obtaining the Standard Model
region.}}
\label{final3}
\end{tighten}
\end{figure}

\begin{figure}
\psfig{file=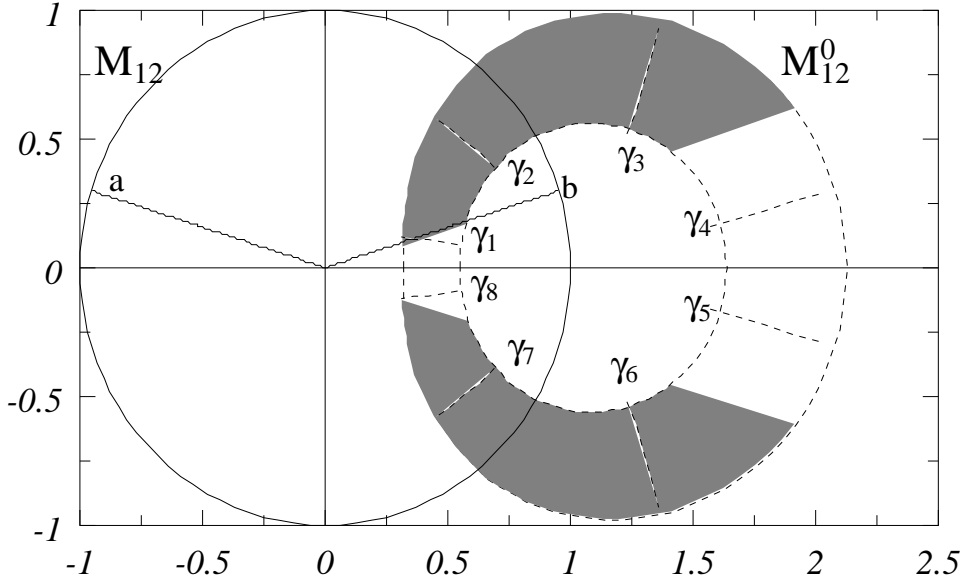,width=5.0in}
\vskip 0.2in
\begin{tighten}
\caption[a]{\small{
The complex $M_{12}$ plane in units of $\Delta m_B/2$.
The points $a$ and $b$ and the curves $\gamma_1...\gamma_8$ result
from the measurements \apks = 0.3 and \app = -0.7.
The shaded region corresponds to the allowed Standard Model parameter
space coming from a measurement of $a_{SL}<0$.
We have used $\sqrt{B_B}f_B= 200$ MeV in obtaining the Standard Model
region.}}
\label{final1}
\end{tighten}
\end{figure}

\begin{figure}
\psfig{file=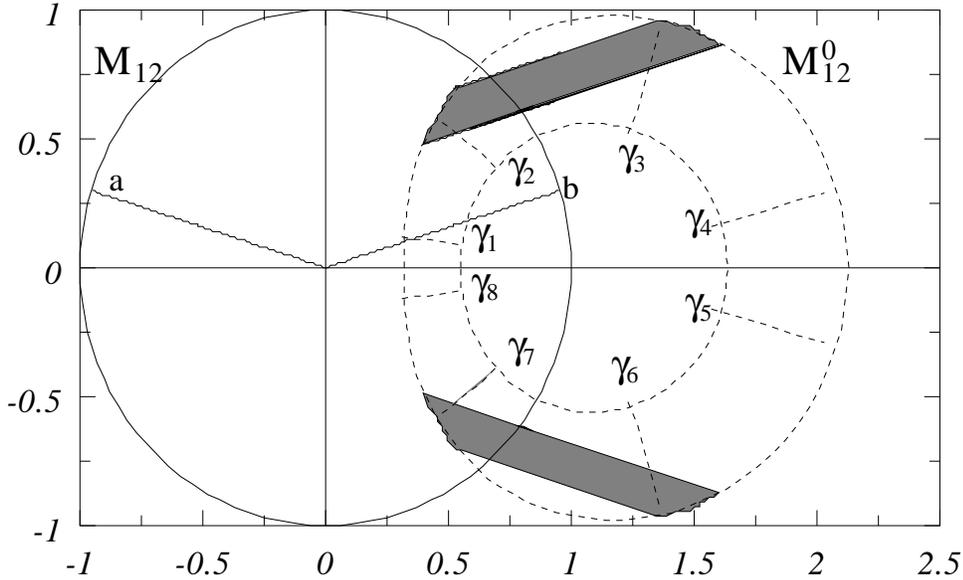,width=5.0in}
\vskip 0.2in
\begin{tighten}
\caption[a]{\small{
The complex $M_{12}$ plane in units of $\Delta m_B/2$.
The points $a$ and $b$ and the curves $\gamma_1...\gamma_8$ result
from the measurements \apks = 0.3 and \app = -0.7.
The shaded region corresponds to the allowed Standard Model parameter
space coming from a measurement of $a_{SL}=(-5\pm 1)\times 10^{-3}$.
We have used $\sqrt{B_B}f_B= 200$ MeV in obtaining the Standard Model
region.}}
\label{final2}
\end{tighten}
\end{figure}

In Figs. \ref{final3}, \ref{final1}, and \ref{final2} we 
use three hypothetical scenarios to highlight the effects 
of combining \apks\ and \app\ 
with \asl\ in constraining the allowed \sm\ parameter space. 
As before, we use \fb\ = 200 MeV and $0.06 \le a \le 0.10$ 
to construct the allowed \sm\  region, and assume that 
$a_{\psi K_S} = 0.3$, and $a_{\pi\pi}=-0.7$
have been measured. In all three figures, 
the points labeled ${a}$ and ${b}$ correspond to the
two-fold ambiguity in obtaining the phase of \moff\
and $\gamma_1 ... \gamma_8$ represent the eight-fold ambiguity in
obtaining $M_{12}^0$.

We first discuss what the constraint 
$|a_{SL}| < 5 \times 10^{-3}$ would teach us. 
This should be acheivable in one years running at the asymmetric 
$B$ factories \cite{Yamamoto}. 
As can be seen from Fig. \ref{triangle}, $M_{12}^0$ must lie in a band 
of width $a_{SL}/(0.8\times 10^{-2})$ above or below $M_{12}$. We
illustrate this in Fig. ~\ref{final3} where we can see that some of 
the allowed parameter space is ruled out by this constraint.

Next, in Fig. ~\ref{final1},
we illustrate what a measurement of $a_{SL}<0$ would teach us. 
From Eq. (\ref{asl}) we see that $a_{SL} < 0$ implies
$-\pi < 2\theta < 0$, thus $M_{12}^0$ must be either above the
$M_{12}$ vector labeled $b$, or below the one labeled $a$.
This corresponds to the shaded region in the figure, where we see 
four of the allowed $\gamma$ curves for $M_{12}^0$ have been ruled out.
The fact that one can obtain this significant restriction on
the \sm\ allowed region just from the sign of \asl\ has the major
advantage that one does not need a very
precise measurement of \asl, just one that is $3\sigma$ from
zero. This would be useful if \asl\ turns out to be large.

Finally, in Fig. \ref{final2} 
we show the constraints for the same value of \apks\ and
\app, but now with a measurement of $a_{SL} = (-5 \pm 1) \times
10^{-3}$. In this case the \sm\ point must lie in one of the two
shaded bands parallel to the \moff\ vectors $a$ and $b$ respectively. 
The
width of the bands includes both the assumed experimental error in the 
measurement of \asl, as well as the theoretical uncertainty
in the coefficient of \asl\ [c.f Eq. (\ref{asl})].
Notice that for particular values of $\gamma$ we now know both $\sin
2\theta$ and $r^2$, hence one has not only resolved the \sm\ parameters, but
also the new physics ones.

A crucial ingredient in the discussion so far is the reliability of
the \sm\ calculation of $\Gamma_{12}^0$, which is essentially a long
distance quantity. The calculation 
consists of an inclusive sum over final states that are common to the
$B$ and the $\bar B$. Thus, it is reliable to the extent that one can
use the notion of local quark-hadron duality in doing such a
calculation. Although this is expected to be correct \cite{LM},
there have been objections to this calculation \cite{BjW,Wolfenstein}, and
it is important to be able to test its accuracy.
One such test is by measuring CP violation in semi-inclusive hadronic $B$
decays as proposed in \cite{BBD2}. The $B$ decays into these
semi-inclusive channels are precisely those that are summed to give
$\Delta \Gamma$ and \asl. An agreement between the measurements and
theoretical expectation would support the use of local quark-hadron duality in
this calculation. 
Another test is available within the $B_s$ system, 
where there exist two complementary calculations of the quantity
$\Delta\Gamma = Re \Gamma_{12}$. 
One done
by actually summing over common final states \cite{Aleksan} and one
using quark-hadron duality \cite{BBD}. The fact that both give similar 
answers and with the same sign could be an indication of the
reliability of the quark level calculation. 
More importantly, $\Delta\Gamma_s$ may actually be large enough to be
measurable. A measurement of 
$\Delta\Gamma_s$ which agrees with the quark level prediction would be 
further indication of the correctness of the calculation.


To conclude, we have discussed the information one can obtain from a
measurement of \asl. The \sm\ prediction $a_{SL} \lesssim 10^{-3}$ is
robust within the assumption of quark-hadron duality. Thus, a
measurement in contradiction with this limit would indicate the
presence of physics beyond the \sm. Within this theoretical framework, 
we have shown how a measurement of \asl\ could help constrain the CKM
parameters, and remove some of the discrete ambiguities in their
phases. This method depends on the ratio of bag factors $B_S/B_B$
which is poorly known at present, but for which there should be
improved lattice calculations available soon. 
Finally, we have presented the analysis in terms of \moff, which 
affords us something quite 
beyond what is available with the unitarity triangle.
This graphical representation \cite{GOTO} correctly represents
where the real uncertainties lie.  They are
not in $\Delta m_B$, which is known quite well, but in our estimation
of the Standard Model prediction of \mot.

\newpage

{\em Acknowledgements} The authors would like to thank 
M. Beneke, G. Buchalla, Y. Grossman, L. Randall
and L. Wolfenstein for useful discussions. This work was
supported in part by the National Science Foundation under grant 
PHY-95-147947.

\end{document}